\documentclass{emulateapj}
\usepackage{graphicx}

\begin{document}

\title{Can the maximum mass of neutron stars rule out any equation of state of dense stellar matter before gravity is well understood?}

\author{De-Hua Wen$^{1,2}$, Bao-An Li\footnote{Corresponding author, Bao-An\_Li$@$Tamu-Commerce.edu} $^{2,3}$, Lie-Wen Chen$^4$}

\affil{$^1$Department of Physics, South China University of
Technology, Guangzhou 510641, P.R. China} \affil{$^2$ Department
of Physics and Astronomy, Texas A\&M University-Commerce,
Commerce, Texas 75429-3011, USA} \affil{$^3$ Department of Applied
Physics, Xi'an Jiao Tong University, Xi'an 710049, P.R. China}
\affil{$^4$ Department of Physics, Shanghai Jiao Tong University,
Shanghai 200240, P.R. China}

\begin{abstract}
Probably No! As an example, using soft EOSs consistent with
existing terrestrial nuclear laboratory experiments for hybrid
neutron stars containing a quark core described with MIT bag model
using reasonable parameters, we show that the recently discovered
new holder of neutron star maximum mass PSR J1614-2230 of
$1.97\pm0.04M_{\odot}$ can be well described by incorporating a
Yukawa gravitational correction that is consistent with existing
constraints from neutron-proton and neutron-lead scatterings as
well as the spectroscopy of antiproton atoms.
\end{abstract}


\section{Introduction}

What is gravity? Are there additional spacetime dimensions? These
are among the Eleven Science Questions for the New Century
identified by the Committee on the Physics of the Universe, US
National Research Council \citep{11questions}. Interestingly,
despite the fact that gravity is the first force discovered in
nature, the quest to unify it with other fundamental forces
remains elusive because of its apparent weakness at
short-distance, see, e.g., refs.
\citep{Ark98,Pea01,Hoy03,Long03,Jean03,Boehm04a,Boe04,Dec05}. In
developing grand unification theories, the conventional
inverse-square-law (ISL) of Newtonian gravitational force has to
be modified due to either the geometrical effect of the extra
spacetime dimensions predicted by string theories and/or the
exchange of weakly interacting bosons, such as the neutral spin-1
vector $U$-boson \citep{Fayet}, proposed in the super-symmetric
extension of the Standard Model, see, e.g., refs.
\citep{Adel03,Adel09,Fis99,New09,Uzan03,Rey05} for recent reviews.
The modified gravity has also been proposed as an explanation for
the present period of cosmological acceleration, see, e.g., ref.\
\citep{Ded08}. The search for evidence of modified gravity is at
the forefront of research in several sub-fields of natural
sciences including geophysics, nuclear and particle physics, as
well as astrophysics and cosmology,  see, e.g., refs.\
\citep{Fujii71,Pea01,Hoy03,Ark98,Long03,Adel09,Kap07,Nes08,Kam08,Aza08,Ger10,Luc10}.
Various upper limits on the deviation from the ISL has been put
forward down to femtometer range. Since the composition of neutron
stars are determined mainly by the weak and electromagnetic forces
through the $\beta$ equilibrium and charge neutrality conditions
while their stability is maintained by the balance of strong and
gravitational forces, neutron stars are thus a natural testing
ground of grand unification theories of fundamental forces.
Moreover, neutron stars are among the densest objects with the
strongest gravity in the Universe, making them ideal places to
test strong-field predictions of General Relativity (GR)
\citep{Psa08}. The masses and radii of neutron stars are solely
determined by both the strong-field behavior of gravity and the
Equation of State (EOS) of dense stellar matter. However, there is
no fundamental reason to choose Einstein's equations over other
alternatives and it is known that the GR theory itself may break
down at the limit of very strong gravitational fields, see, e.g.,
ref. \citep{Psa08} for a comprehensive review. In fact, effects of
modified gravity on properties of neutron stars have been under
intense investigation. As expected, results of these studies are
strongly model dependent, see, e.g., refs.
\citep{Ger01,Wis02,Aza08,Kri09,Wen0911}. Nevertheless, it is very
interesting to note that alternative gravity theories that have
all passed low-field tests but diverge from GR in the strong-field
regime predict neutron stars with significantly different
properties than their GR counterparts \citep{Ded03}. Moreover, the
deviations for neutron star properties from the GR predictions for
these theories are larger than the uncertainty due to the poorly
known EOS of dense matter in neutron stars. It was also clearly
shown that the neutron star maximum mass alone can not distinguish
gravity theories \citep{Ded03}. Furthermore, in the endeavor of
testing GR theory of gravity using properties of neutron stars, it
is known that there is a degeneracy between the matter content and
gravity. This degeneracy is tied to the fundamental Strong
Equivalence Principle and can only be broken by using at least two
independent observables \citep{Yun10}.

Recently, using the general relativistic Shapiro delay the mass of
PSR J1614-2230 was precisely measured to be
$1.97\pm0.04M_{\odot}$\citep{Demo10}, making it the new holder of
the maximum mass of neutron stars. Comparing with mass-radius
relations predicted from solving the TOV equation using various
EOSs within GR theory of gravity, it was shown that the mass of
PSR J1614-2230 can rule out almost all soft EOSs especially those
associated with hyperon or boson condensation. While conventional
quark stars with soft EOSs are also ruled out by this observation,
neutron stars with strongly interacting quark cores are allowed
\citep{Demo10,Ozel10,RXu10}. It was further shown that a
transition to quark matter in neutron star cores can occur at
densities comparable to the nuclear saturation density $\rho_0$
only if the quarks are strongly interacting and are color
superconducting \citep{Ozel10}. The mass of PSR J1614-2230 was
then used to constrain the interacting parameters of quarks. It
was also shown that neutron stars with interacting quark clusters
in their cores or solid quark stars can be very massive. Using the
Lennard-Jones potential for interactions between quark clusters,
the mass of the PSR J1614-2230 was used to constrain the number of
quarks inside individual quark clusters \citep{RXu10}. In this
work, using soft nuclear EOSs for hybrid stars containing a quark
core described by the MIT bag model with reasonable parameters, we
show that the mass of PSR J1614-2230 is readily obtained by
incorporating a Yukawa gravitational correction that is consistent
with existing constraints from terrestrial nuclear laboratory
experiments.

\section{Non-Newtonian gravity and model EOS of hybrid stars}
Fujii \citep{Fujii71} first proposed that the non-Newtonian
gravity can be described by adding a Yukawa term to the
conventional gravitational potential between two objects of mass
$m_1$ and $m_2$, i.e.,
\begin{equation}
V(r)=-\frac{Gm_{1}m_{2}}{r}(1+\alpha e^{-r/\lambda}),
\end{equation}
where $\alpha$ is a dimensionless strength parameter, $\lambda$ is
the length scale and $G$ is the  gravitational constant. In the
boson exchange picture, $\alpha=\pm g^2/(4\pi Gm_b^{2})$ where
$\pm$ stands for scalar/vector bosons and $\lambda=1/\mu$ (in
natural units). The $g^2$ and $\mu$ are the boson-baryon coupling
constant and the boson mass, respectively. The light and weakly
interacting $U$-boson is a favorite candidate mediating the extra
interaction \citep{Fayet,Kri09,Zhu07}. Similar to the degeneracy
between matter content and gravity, there appears to be a duality
of incorporating effects of the Yukawa term in either the TOV
equation or the input EOS. Nevertheless, according to Fujii
\citep{Fuj2}, the Yukawa term is simply part of the matter system
in general relativity. Therefore, only the EOS is modified and the
TOV equation remains the same. Within the mean-field
approximation, the extra energy density due to the Yukawa term is
\citep{Long03,Kri09}
\begin{equation}\label{EDUB}
\varepsilon_ {_{\textrm{\scriptsize{UB}}}}= \frac{1}{2V}\int
\rho(\vec{x}_{1})\frac{g^{2}}{4\pi}\frac{e^{-\mu
r}}{r}\rho(\vec{x}_{2})d\vec{x}_{1}d\vec{x}_{2}=\frac{1}{2}\frac{g^{2}}{\mu^{2}}\rho^{2},
\end{equation}
where $V$ is the normalization volume, $\rho$ is the baryon number
density and $r=|\vec{x}_{1}-\vec{x}_{2}|$. Assuming a constant
boson mass independent of the density, one obtains the
corresponding addition to the EOS
$P_{\textrm{\scriptsize{UB}}}=\frac{1}{2}\frac{g^{2}}{\mu^{2}}\rho^{2}$,
which is just equal to the additional energy density. As it was
emphasized by Fujii \citep{Fuj2}, since the new vector boson
contributes to the EOS only through the combination
$g^{2}/\mu^{2}$, while both the coupling constant g and the mass
$\mu$ of the light and weakly interacting bosons are small, the
value of $g^{2}/\mu^{2}$ can be large. On the other hand, by
comparing with the $g^{2}/\mu^{2}$ value of the ordinary vector
boson $\omega$, Krivoruchenko et al. have pointed out that as long
as the $g^{2}/\mu^{2}$ value of the U boson is less than about 200
GeV$^{-2}$ the internal structures of both finite nuclei and
neutron stars will not change \citep{Kri09}. However, global
properties of neutron stars can be significantly modified
\citep{Kri09,Wen0911} One of the key characteristics of the Yukawa
correction is its composition dependence, unlike Einstein's
gravity. Thus, ideally one needs to use different coupling
constants for various baryons existing in neutron stars. Moreover,
to our best knowledge, it is unknown if there is any and what
might be the form and strength of the Yukawa term in the
hadron-quark mixed phase and the pure quark phase. Nevertheless,
instead of introducing more parameters, for the purpose of this
exploratory study, we assume that the
$P_{\textrm{\scriptsize{UB}}}$ term is an effective Yukawa
correction existing in all phases with the $g$ considered as an
averaged coupling constant.
\begin{figure}
\includegraphics[width=0.5\textwidth]{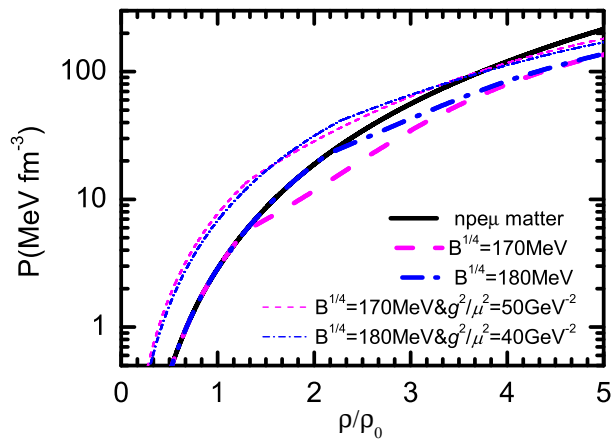}
\vspace{-0.7 cm} \caption{\label{fig1}  (Color online) Model EOSs
for hybrid stars with and without the Yukawa contribution using
MIT bag constant $B^{1/4}$=170 MeV and $B^{1/4}$=180 MeV,
respectively.}
\end{figure}

Including the Yukawa term the EOS becomes $P=P_
{\textrm{\scriptsize{0}}}+P {_{\textrm{\scriptsize{UB}}}}$ where
$P_ {0}$ is the conventional pressure inside neutron stars. For
the latter, we use typical model EOSs for hybrid stars containing
a quark core covered by hyperons and leptons. The quark matter is
described by the MIT bag model with reasonable parameters widely
used in the literature \citep{MIT1,MIT2}. The hyperonic EOS is
modelled by using an extended isospin- and momentum-dependent
effective interaction (MDI \citep{Das03}) for the baryon octet
with parameters constrained by empirical properties of symmetric
nuclear matter, hyper-atoms, and heavy-ion reactions \citep{xu10}.
In particular, the underlying EOS of symmetric nuclear matter is
constrained by comparing transport model predictions with data on
collective flow and kaon production in relativistic heavy-ion
collisions \citep{Pawel,LCK}. Moreover, the nuclear symmetry
energy $E_{sym}(\rho)$ with this interaction is chosen to increase
approximately linearly with density (i.e., the MDI interaction
with a symmetry energy parameter x=0 \citep{xu10}) in agreement
with available constraints around and below the saturation density
\citep{Cxu10}. It is worth noting that this $E_{sym}(\rho)$ is
very similar to the well-known APR prediction up to about
$5\rho_{0}$ \citep{APR,Xiao09}. The Gibbs construction was adopted
to describe the hadron-quark phase transition \citep{Glen01}.
Similar to the previous work in the literature, the hybrid star is
divided into the liquid core, inner crust and outer crust from the
center to surface. For the inner crust, a parameterized EOS of
$P_{\textrm{\scriptsize{0}}}=a+b \epsilon^{4/3}$ is adopted as in
refs.\ \citep{XCLM09b}. For the outer crust, the BPS
EOS~\citep{BPS} is adopted. As an example, shown in
Fig.~\ref{fig1} are the EOSs for hybrid stars with MIT bag
constant $B^{1/4}=170$ MeV and $B^{1/4}=180$ MeV, respectively,
with and without the Yukawa contribution. The corresponding
hadron-quark mixed phase above/below the pure hadron/quark phase
covers the density range of $\rho/\rho_0=1.31$ to 6.56 and
$\rho/\rho_0=2.19$ to 8.63, respectively. Including the Yukawa
term the EOS stiffens as the value of $g^{2}/\mu^{2}$ increases.
It is seen that the two sets of parameters, $B^{1/4}=170$ MeV and
$g^{2}/\mu^{2}=$ 50~GeV$^{-2}$ or $B^{1/4}=180$ MeV and
$g^{2}/\mu^{2}=$ 40~GeV$^{-2}$, lead to approximately the same
total pressure. As a reference, the MDI EOS for the $npe\mu$
matter is also shown.

\section{Maximum mass of hybrid stars with non-Newtonian gravity}
\begin{figure}
\includegraphics[width=0.5\textwidth]{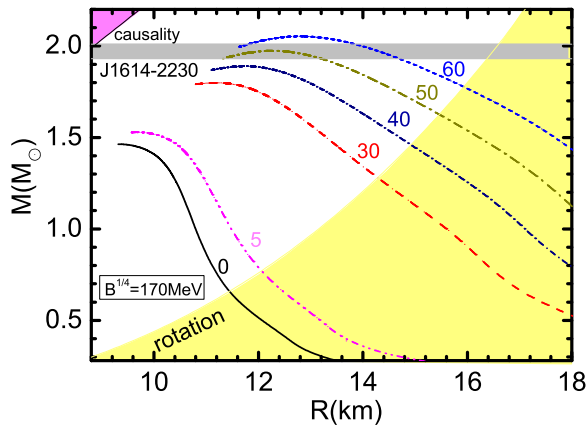}
\vspace{-0.7 cm} \caption{\label{fig2}  (Color online) The
mass-radius relation of static neutron stars with $B^{1/4}$=170
MeV and various values of $g^{2}/\mu^{2}$ in units of GeV$^{-2}$
indicted using numbers above the lines.}
\end{figure}
\begin{figure}
\includegraphics[width=0.5\textwidth]{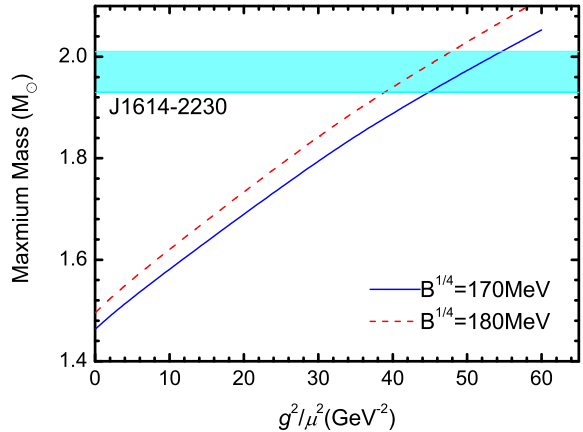}
\vspace{-0.7 cm} \caption{\label{fig3}  (Color online) The maximum
mass of neutron stars as a function of $g^{2}/\mu^{2}$ with
$B^{1/4}$=170 MeV and $B^{1/4}$=180 MeV, respectively.}
\end{figure}
As an example, shown in Fig. \ref{fig2} is the mass-radius
relation of hybrid stars with the bag constant $B^{1/4}$=170 MeV
and varying values of $g^{2}/\mu^{2}$. First of all, without the
Yukawa contribution (black solid line) the maximum stellar mass
supported is only about $1.46 M_{\odot}$. Including the Yukawa
term, as the EOSs are increasingly stiffened with larger values of
$g^{2}/\mu^{2}$, the maximum stellar mass increases. With
$g^{2}/\mu^{2}=50$ GeV$^{-2}$ the maximum mass of $1.97 M_{\odot}$
is just in the middle of the measured mass band of PSR J1614-2230.
The corresponding radius is about 12.4 km. To see more clearly
relative effects of the bag constant B and the Yukawa term, shown
in Fig.~\ref{fig3} are the maximum stellar masses as a function of
$g^{2}/\mu^{2}$ with $B^{1/4}=170$ MeV and $B^{1/4}=180$ MeV,
respectively. As expected, with $B^{1/4}=180$ MeV a smaller value
of $g^{2}/\mu^{2}=40~$GeV$^{-2}$ is needed to obtain a maximum
mass consistent with the observed mass of PSR J1614-2230.

\section{Comparison with terrestrial constraints on non-Newtonian gravity at short distance}
\begin{figure}
\includegraphics[width=0.5\textwidth]{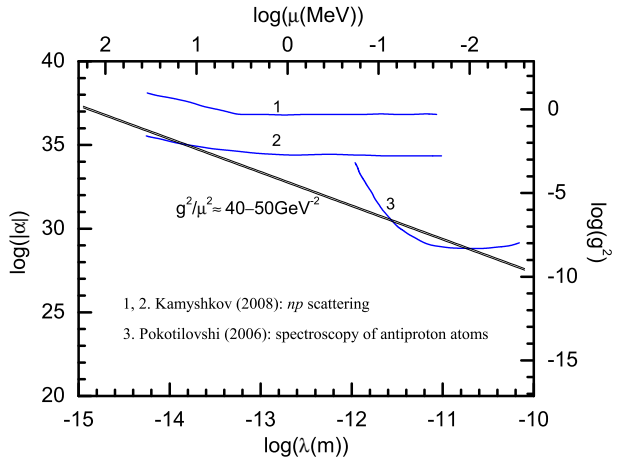}
\vspace{-0.7 cm} \caption{\label{fig4}  (Color online) Constraints
on the strength and range of the Yukawa term from terrestrial
nuclear experiments in comparison with $g^{2}/\mu^{2}\approx 40-50
GeV^{-2}$.}
\end{figure}
As mentioned earlier, significant efforts have been devoted to
constrain the possible non-Newtonian gravity using terrestrial
experiments. These experiments have established a clear trend of
increased strength $\alpha$ at shorter length $\lambda$. In the
short range down to $\lambda \approx 10^{-14}-10^{-8}$ m,
neutron-proton and neutron-lead scattering data as well as the
spectroscopy of antiproton atoms have been used to set upper
limits on the value of $g^{2}/\mu^{2}$ (or equivalently the
$|\alpha|~ vs ~\lambda$). It is thus interesting to compare the
values of $g^{2}/\mu^{2}$ necessary to support the PSR J1614-2230
within the model presented above with the constraints extracted
from terrestrial experiments. While the range parameter $\lambda$
is expected to be much larger (smaller) than the radii of finite
nuclei (neutron stars), the maximum mass of neutron stars alone is
not sufficient to set separate constraints on the values of
$\alpha$ and $\lambda$. Shown in Fig. \ref{fig4} is a comparison
with the terrestrial constraints \citep{Kam08,BARB75,POKO06,Nes08}
in the $|\alpha|~ vs ~\lambda$ plane. The straight line is for
$g^{2}/\mu^{2}\approx 40-50$~GeV$^{-2}$. It is seen that the
values of $g^{2}/\mu^{2}$ necessary to describe the maximum mass
of PSR J1614-2230 are consistent with the upper limits from the
terrestrial experiments.

\section{Conclusions}
Among all fundamental forces, gravity remains the most uncertain
one despite being the first discovered in nature. Neutron stars
are natural testing grounds of grand unification theories of
fundamental forces. In particular, they are ideal places to test
GR predictions at the strong-field limit. Interpretations of
observed properties of neutron stars require a comprehensive
understanding of both gravity and the EOS of dense stellar matter.
Before strong-field gravity is well understood, it is unlikely
that the maximum mass of neutron stars alone can rule out any EOS.
As an example, using soft nuclear EOSs consistent with existing
terrestrial experiments for hybrid stars containing a quark core
described by the MIT bag model with reasonable parameters, the
maximum mass of PSR J1614-2230 is readily obtained by
incorporating the Yukawa gravitational correction that is
consistent with existing constraints from terrestrial nuclear
laboratory experiments.

We thank W.Z. Jiang, W. G. Newton, A.W. Steiner and Y. Zhang for
useful discussions. D.H. Wen is supported in part by the National
Natural Science Foundation of China under Grant No.10947023 and
the Fundamental Research Funds for the Central University, China
under Grant No.2009ZM0193. B.A. Li is supported in part by the US
National Science Foundation under grant PHY-0757839, the National
Aeronautics and Space Administration under grant NNX11AC41G issued
through the Science Mission Directorate and the Texas Coordinating
Board of Higher Education under grant No. 003565-0004-2007. L.W.
Chen is supported in part by the National Natural Science
Foundation of China under Grant Nos. 10675082 and 10975097, MOE of
China under project NCET-05-0392, Shanghai Rising-Star Program
under Grant No. 06QA14024, the SRF for ROCS, SEM of China, the
National Basic Research Program of China (973 Program) under
Contract No. 2007CB815004.


\clearpage


\begin{thebibliography}{99}
\bibitem[Adelberger et al.(2003)] {Adel03} Adelberger, E. G.,  Heckel, B. R., \&  Nelson, A. E. 2003, Annu. Rev. Nucl. Part. Sci., 53, 77
\bibitem[Adelberger et al.(2009)] {Adel09} Adelberger, E. G.,   Gundlach, J. H.,  Heckel, B. R.,  Hoedl, S.,\&  Schlamminger S. 2009, Prog. Part. Nucl. Phys., 62, 102
\bibitem[Akmal et al.(1998)] {APR} Akmal, A., Pandharipande, V. R. and Ravenhall, D. G. 1998, Phys. Rev. c, 58, 1804
\bibitem[Arkani-Hamed et al.(1998)]{Ark98}  Arkani-Hamed, N.,   Dimopoulos, S., \&  Dvali, G. 1998, Phys Lett. B, 429, 263; \prd, 59, 086004
\bibitem[Azam  et al.(2008)]{Aza08}  Azam, M., Sami, M., Unnikrishnan,  C. S., \&  Shiromizu, T. 2008, Phys. Rev. D, 77, 101101
\bibitem[Barbieri et al.(1975)]{BARB75} Barbieri,  R., \&
Ericson, T. 1975, Phys. Lett. B, 57, 270
\bibitem[Baym et al.(1971)] {BPS}  Baym, G.,  Pethick, C., \&  Sutherland,  P. 1971, \apj, 170, 299
\bibitem[Boehm et al.(2004a)] {Boehm04a} Boehm, C.,   Hooper, D.,  Silk, J., Casse, M. \&  Paul, J. 2004, Phys. Rev. Lett., 92, 101301
\bibitem[Boehm et al.(2004b)]{Boe04}Boehm, C., \& Fayet, P. 2004, Nucl. Phys. B, 683, 291
\bibitem[Chodos et al.(1974)] {MIT1} Chodos, A., Jaffe, R. L., Johnson, K., Thorn,  C. B., \&  Weisskopf, V. F. 1974, Phys. Rev. D, 9, 12
\bibitem[CPUNRC(2003)]{11questions}Committee on the Physics of the Universe, National Research Council 2003, Connecting Quarks with the Cosmos:Eleven Science Questions for the New Century (The National academies Press) 
\bibitem[Danielewicz et al.(2000)]{Pawel} Danielewicz, P., Lacey, R., \& Lynch, W. G. 2000, Science, 298, 1592
\bibitem[Das et al.(2003)] {Das03}  Das, C.B.,  Das Gupta, S.,  Gale, C., \&  Li, B. A. 2003, \prd, 67, 034611
\bibitem[Decca et al.(2005)] {Dec05}  Decca, R.S.,L\'opez,  D., Chan, H. B., Fischbach, E., Krause, D. E., \&  Jamell, C. R. 2005, \prl 94, 240401
\bibitem[DeDeo et al.(2003)] {Ded03}  DeDeo, S., \&  Psaltis, D. 2003, \prl, 90,  141101
\bibitem[DeDeo et al.(2008)] {Ded08}  DeDeo, S., \&  Psaltis, D. 2008, \prd, 78, 064013
\bibitem[{Demorest et al.(2010)}]{Demo10} Demorest, P., Pennucci, T., Ransom, S., Roberts, M., \& Hessels, J.\ 2010, \nat, 467, 1081
\bibitem[Fayet(2009)] {Fayet}  Fayet, P. 2009, Phys. Lett. B, 675, 267
\bibitem[Fischbach(1999)] {Fis99}  Fischbach, E., \&  Talmadge, C. L. 1999, The Search for Non-Newtonian Gravity(Springer-Verlag, New York, Inc. ) 
\bibitem[Fujii(1971)] {Fujii71}  Fujii, Y. 1971, \nat, 234, 5
\bibitem[Fujii(1988)] {Fuj2} Fujii, Y. 1988,  in Large Scale Structures of the Universe, page 471-477 (Eds. J. Audouze et al., International Astronomical Union.)
\bibitem[Geraci et al.(2010)]{Ger10} Geraci, A. A.,  Papp, S. B., \&  Kitching, J.  2010, \prl, 105, 101101
\bibitem[Germani et al.(2001)] {Ger01}  Germani, C., \& Maartens, R. 2001, \prd, 64, 124010
\bibitem[Glendenning(2001)]{Glen01} Glendenning,  N.K. 2001, Phys. Rep., 342,  393
\bibitem[Heinz et al.(1986)] {MIT2} Heinz, U., Subramanian,   P. R.,  Stocker, H., \& Greiner, W. 1986, Nucl. Phys., 12, 1237
\bibitem[Hoyle(2003)]{Hoy03} Hoyle, C. D. 2003, \nat, 421, 899
\bibitem[Jean et al.(2003)]{Jean03} Jean, P., et al. 2003, A\&A, 407, L55
\bibitem[Kamyshkov et al.(2008)]{Kam08} Kamyshkov, Y.,  Tithof J., \&  Vysotsky, M. 2008, Phys. Rev. D, 78, 114029
\bibitem[Kapner et al.(2007)] {Kap07} Kapner, D.J., Cook, T. S., Adelberger, E. G., Gundlach, J. H., Heckel, B. R., Hoyle, C. D., \&  Swanson, H. E. 2007, \prl, 98, 021101
\bibitem[Krivoruchenko et al.,(2009)] {Kri09} Krivoruchenko, M.I., Simkovic, F., \&  Faessler, A.   2009, \prd , 79, 125023
\bibitem[Lai et al.(2010)] {RXu10}  Lai, X. Y., \&  Xu, R. X., 2010, arXiv:1011.0526
\bibitem[Li et al.(2008)] {LCK}  Li, B. A.,  Chen, L. W., \&  Ko, C. M. 2008, Phys. Rep., 464, 113
\bibitem[Long et al.(2003)]{Long03}Long, J. C., et al. 2003, \nat,  421, 922
\bibitem[Lucchesi et al.(2010)]{Luc10}  Lucchesi, D. M., \&  Person R. 2010, \prl, 105, 231103
\bibitem[Nesvizhevsky et al.(2008)] {Nes08} Nesvizhevsky, V. V., et al., 2008, Phys. Rev. D, 77, 034020
\bibitem[Newman(2009)] {New09}  Newman, R. D., Berg, E. C., \& Boynton, P. E. 2009, Space Science Review, 148, 175
\bibitem[\"Ozel et al.(2010)]{Ozel10}\"Ozel, F., Psaltis, D., Ransom, S., Demorest, P., \& Alford, M, 2010, arXiv:1010.5790v1, ApJL in press.
\bibitem[Pea(2001)]{Pea01} Pease, R. 2001, \nat, 411,  986
\bibitem[Pokotilovski(2006)]{POKO06} Pokotilovski, Yu. N. 2006,  Phys. Atom. Nucl., 68, 924
\bibitem[Psaltis(2008)]{Psa08}  Psaltis, D. 2008, Living Reviews in Relativity, 11, 9
\bibitem[Reynaud(2005)] {Rey05} Reynaud, S., \&  Jaekel, M. M. 2005, Int. J. Mod. Phys. 20, 2294
\bibitem[Uzan(2003)] {Uzan03}  Uzan, J. P. 2003, Rev. Mod. Phys., 75, 403
\bibitem[Wen et al.(2009)] {Wen0911} Wen, D. H.,  Li, B. A., \& Chen, L. W., 2009, \prl, 103, 211102
\bibitem[Wiseman(2002)] {Wis02}  Wiseman, T. 2002, \prd, 65, 124007
\bibitem[Xiao et al.(2009)]{Xiao09} Xiao, Z.G., et al., 2009, Phys. Rev. Lett. {\bf 102}, 062502
\bibitem[J. Xu et al.(2010)] {xu10} Xu, J.,  Chen, L. W.,  Ko, C. M., \&  Li, B. A. 2010, \prc,  81, 055803
\bibitem[C. Xu et al.(2010)] {Cxu10}  Xu, C.,  Li, B. A., \&  Chen, L. W. 2010, \prc, 82, 054607
\bibitem[J. Xu et al.(2009)]{XCLM09b} Xu, J., Chen, L. W.,  Li, B. A., \& Ma, H.R. 2009, \apj, 697,  1549
\bibitem[Yunes et al.(2010)] {Yun10}  Yunes, N.,  Psaltis, D.,  Ozel F., \& Loeb, A. 2010, \prd, 81, 064020
\bibitem [Zhu(2007)] {Zhu07}Zhu, S. H. 2007, \prd, 75,  115004
\end{thebibliography}
\end{document}